\documentclass[onecolumn,amsmath,amssymb,superscriptaddress,nofootinbib,11pt,a4paper]{revtex4}
\usepackage{titlesec}
\titleformat{\section}
  {\normalfont\Large\bfseries}  
  {\thesection}                 
  {1em}                         
  {}                            
\titleformat{\subsection}
  {\normalfont\large\bfseries}
  {\thesubsection}
  {1em}
  {}
\pdfoutput=1
\usepackage[T1]{fontenc}
\usepackage{CJK}
\usepackage{xcolor}
\usepackage{amsfonts}
\usepackage{epstopdf}
\usepackage{wrapfig} 
\usepackage{subcaption}
\usepackage{graphicx}  
\usepackage{dcolumn}   
\usepackage{bm}
\usepackage{float}
\usepackage[utf8]{inputenc}
\usepackage{booktabs} 
\usepackage{orcidlink}
\usepackage{multirow}

\begin{document}

\title{Particle decay and energy conservation in the Kerr-Newman black hole}

\author{Bing-Bing Chen }
\affiliation{School of Mathematics, Physics and Statistics, Sichuan Minzu College, \\ Ganzi 626100, China}

\author{Guo-Ping Li }
\email{gpliphys@yeah.net}
\affiliation{ Physics and Astronomy College, China West Normal University,\\ Nanchong 637000, China}

\author{Ke Wang}
\affiliation{School of Material Science and Engineering, Chongqing Jiaotong University, \\Chongqing 400074, China}

\author{Xiao-Xiong Zeng}
\affiliation{College of Physics and Optoelectronic Engineering, Chongqing Normal University, \\Chongqing 401331, China}

\begin{abstract}
{ In this paper, we study the decay of a particle in Kerr--Newman spacetime. Both theoretical analysis and numerical simulations show that when a particle splits in Kerr--Newman spacetime, its mass is inevitably reduced, and this mass deficit is transformed into kinetic energy in the center-of-mass frame. We denote the parent particle as $O$ and the daughter particles as $A$ and $B$. We also find that as the charge parameter $Q$ increases, the specific angular momenta of the parent particle $O$ and the daughter particle $B$ become smaller, the masses of the two daughter particles become larger, the absolute values of their specific energies become smaller, the relative Lorentz factors among the three particles become smaller, and their four-velocities also become smaller. These trends are opposite to those observed when increasing the cosmological constant in Kerr--de Sitter spacetime.

}
\end{abstract}
\maketitle

\newpage
\section{Introduction}
Particle collisions and splitting processes in black hole spacetimes have long been one of the central topics in gravitational theory and high-energy astrophysics, such as the Penrose process \cite{12,13}. By splitting inside the ergosphere, the Penrose process can extract rotational energy from a black hole. In this process, the local energy conservation of particle decay or collision plays a crucial role. In a curved spacetime background, the center-of-mass energy of a two-particle system is determined by the inner product of their four-velocities, and it reduces to the special relativistic form in a local inertial frame \cite{2}, providing a basic framework for examining the energy distribution among the decay products.

For over half a century, the Penrose process has implied a question: what is the relationship between the rest mass of the parent particle before decay and the rest masses of the two daughter particles produced during the process \cite{14}. Classical general relativity textbooks \cite{15} give a detailed description of the Penrose process, but because they assume mass conservation in the process, the description is not entirely precise. It was only recently that Ref.\ \cite{5} discovered in Kerr spacetime that a mass deficit of the parent particle is inevitable, while the energy in the center-of-mass frame of the parent and daughter particles is conserved. Ref.\ \cite{6} extended this to Kerr--de Sitter spacetime, not only proving the energy conservation law in the center-of-mass frame of that spacetime, but also revealing the effects of the decay radius and the cosmological constant on the decay process.

Kerr--Newman spacetime, as the most general stationary axisymmetric black hole solution in Einstein--Maxwell theory \cite{3}, simultaneously contains three parameters---mass, angular momentum, and charge---and its spacetime structure and particle motion are richer. The presence of charge alters the horizon radius, the shape of the ergosphere, and the trajectories of particles \cite{16}, thereby modulating the particle splitting process. Therefore, extending the energy conservation relation of particle decay to the Kerr--Newman black hole is of great significance for understanding the energy extraction mechanism of charged rotating black holes and for testing the universality of energy conservation laws in curved spacetime. It also provides a theoretical reference for studying high-energy particle collisions \cite{17}, gamma-ray bursts \cite{18}, and other astrophysical processes in such backgrounds.

The structure of this paper is as follows: Section~2 studies the characteristics of particle decay and energy conservation in Kerr--Newman black holes; Section~3 presents the numerical calculation results and analysis; and Section~4 provides a summary.

\section{Theoretical study of particle decay and energy conservation in Kerr--Newman black holes}

Before delving into particle decay in Kerr--Newman black holes, we first study the collision or decay process of two particles in a general curved spacetime background. Denote these two particles as $A$ and $B$, with rest masses $m_A$ and $m_B$, respectively. They occupy an identical point in spacetime but possess distinct 4-velocities $U_A^\mu$ and $U_B^\mu$. The corresponding four-momenta can be written as
\begin{equation}
\pi_A^\mu = m_A U_A^\mu,\quad \pi_B^\mu = m_B U_B^\mu. 
\end{equation}
The total four-momentum of the system is then
\begin{equation}
\mathcal{P}^\mu = \pi_A^\mu + \pi_B^\mu = m_A U_A^\mu + m_B U_B^\mu. 
\end{equation}
The center-of-mass energy $E_{\rm cm}$ of this two-particle system is \cite{1}
\begin{equation}
E_{\rm cm}^{2} = -\mathcal{P}^\mu \mathcal{P}_\mu = m_A^{2} + m_B^{2} - 2m_A m_B g_{\mu\nu}U_A^\mu U_B^\nu. 
\end{equation}

We next consider a local inertial frame where particle $A$ is at rest. In this reference frame, the four-velocities of the two particles can be taken as
\begin{equation}
U_A^\mu = (1,0,0,0), \quad U_B^\mu = (\gamma, \gamma \vec{w}),
\end{equation}
where $\gamma = 1/\sqrt{1-w^{2}}$ denotes the relative Lorentz factor between the two particles, and $w = |\vec{w}|$ is the relative speed. According to the equivalence principle, the center-of-mass energy in this local inertial frame obeys the special relativistic relation \cite{2}
\begin{equation}
E_{\rm cm}^{2} = m_A^{2} + m_B^{2} + 2m_A m_B \gamma. 
\end{equation}
In the above expressions, the formula for the center-of-mass energy appears twice. By equating these two expressions for the center-of-mass energy, we obtain
\begin{equation}
-\,g_{\mu\nu}U_A^\mu U_B^\nu = \gamma = 1/\sqrt{1-w^{2}}. \label{6}
\end{equation}

We now turn to the Kerr--Newman black hole. In the Kerr--Newman black hole background, a parent particle $O$ with rest mass $m_{O}$ decays into particles $A$ and $B$. We consider the parent particle to be uncharged and also assume that the daughter particles after splitting are uncharged. The conservation of four-momentum requires that the coordinate components satisfy
\begin{equation}
\pi_{t O} = \pi_{t A} + \pi_{t B},\quad \pi_{r O} = \pi_{r A} + \pi_{r B},\quad \pi_{\theta O} = \pi_{\theta A} + \pi_{\theta B},\quad \pi_{\phi O} = \pi_{\phi A} + \pi_{\phi B}.\label{7}
\end{equation}
Here $\pi_{\mu\,i}$ denotes the $\mu$-component of the four-momentum of particle $i\;(i=O,A,B)$. The energy conservation relation we shall prove in Kerr--Newman spacetime is
\begin{equation}
m_O = m_A\,\gamma_{O A} + m_B\,\gamma_{O B}, \label{8}
\end{equation}
where $\gamma_{O i} = 1/\!\sqrt{1 - w_{O i}^{2}}$ represents the relative Lorentz factor between the parent particle $O$ and the daughter particle $i$. If this relation holds, then when $\gamma_{O A}, \gamma_{O B} > 1$ (i.e., the decay products have relative velocities and separate from each other), we must have $m_A + m_B < m_O$, which means that a mass deficit exists during the decay process, and the mass deficit is transformed into kinetic energy in the center-of-mass frame. Using Eq.~\eqref{6}, the right-hand side of Eq.~\eqref{8} can be represented as
\begin{equation}
m_A \gamma_{O A} + m_B \gamma_{O B} = -\,g_{\mu \nu}\left( m_A U_O^{\mu} U_A^{\nu} + m_B U_O^{\mu} U_B^{\nu} \right), \label{9}
\end{equation}
where $U_i^{\mu}$ stands for the four-velocity of particle $i$. In what follows, we will prove that the right-hand side of Eq.~\eqref{9} is exactly equal to $m_O$.

We use the Boyer--Lindquist coordinate system $(t, r, \theta, \phi)$ with the signature $(-, +, +, +)$, and the Kerr--Newman metric is given by \cite{3}
\begin{equation}
ds^2 = -\frac{\Delta}{\rho^2} ( dt - a \sin^2\theta  d\phi )^2 + \frac{\rho^2}{\Delta} dr^2 + \rho^2 d\theta^2 \
+ \frac{\sin^2\theta}{\rho^2} \left[ (r^2 + a^2) d\phi - a  dt \right]^2,
\end{equation}
where
\begin{equation}
\Delta \equiv r^2 - 2Mr + a^2 + Q^2, \quad 
\rho^2 \equiv r^2 + a^2 \cos^2\!\theta.
\end{equation}
The four-velocity $U_i^\mu$ of particle $i$ can be computed by solving the Hamilton--Jacobi equation using the method of separation of variables \cite{16,4}
\begin{equation}
\begin{aligned}
\rho^2 U_i^t   &= -a\,(aE_i \sin^2 \theta - L_i) + \frac{(r^2 + a^2)T_i}{\Delta},\\
\rho^2 U_i^r   &= \pm \sqrt{R_{i}}, \\
\rho^2 U_i^\theta &= \pm \sqrt{\Theta_{i}}, \\
\rho^2 U_i^\phi &= -\Bigl(aE_i - \frac{L_i}{\sin^2 \theta}\Bigr) + \frac{aT_i}{\Delta},
\end{aligned}
\end{equation}
where
\begin{equation}
\begin{aligned}
T_i &= E_i(r^2 + a^2) - L_i a , \\
R_{i} &= T_i^2 - \Delta \left[ r^2 + (L_i - aE_i)^2 + \mathcal{Q}_i \right] , \\
\Theta_{i} &= \mathcal{Q}_i - \cos^2 \theta \left[ a^2 (1 - E_i^2) + \frac{L_i^2}{\sin^2 \theta} \right] .
\end{aligned}
\end{equation}
Here $E_i$ and $L_i$ are the specific energy and the axial component of the specific angular momentum of particle $i$, and $\mathcal{Q}_i$ is its Carter constant.
Note that the decay products separate from each other, so the radial components of their four-velocities $U_A^r$ and $U_B^r$ have opposite signs, and similarly for the angular components $U_A^\theta$ and $U_B^\theta$. The four-momentum of the particle is given by
\begin{equation}
\pi_{\mu i} = m_i  g_{\mu\nu} U_i^\nu .
\end{equation}
Accordingly, the coordinate components can be written as
\begin{equation}
\pi_{t i} = -m_i E_i,\quad
\pi_{r i} = \pm m_i \frac{\sqrt{R_{i}}}{\Delta},\quad
\pi_{ \theta i} = \pm m_i \sqrt{\Theta_{i}},\quad
\pi_{\phi i} = m_i L_i .\label{15}
\end{equation}
Similarly, the radial components $\pi_{r A}$ and $\pi_{r B}$ have opposite signs, and the angular components $\pi_{\theta A}$ and $\pi_{\theta B}$ have opposite signs as well. Now, by convention, we take $U_A^r$ and $U_A^\theta$ to be negative, and $U_B^r$ and $U_B^\theta$ to be positive. Then, expanding the right-hand side of Eq.~\eqref{9} yields
\begin{equation}
\begin{aligned}
&-\,g_{\mu \nu}\left( m_A U_O^{\mu} U_A^{\nu} + m_B U_O^{\mu} U_B^{\nu} \right)=\\
&\frac{1}{\Delta^2 \rho^6} \Bigl[ 
\frac{1}{4} \Delta \sqrt{R_O} \, (m_A \sqrt{R_A} - m_B \sqrt{R_B}) \, (a^2 + 2r^2 + a^2 \cos 2\theta)^2 \\
&\quad + \frac{1}{4} \Delta^2 (a^2 + 2r^2 + a^2 \cos 2\theta)^2 \sqrt{\Theta_O} \, (m_A \sqrt{\Theta_A} - m_B \sqrt{\Theta_B}) \\
&\quad - \bigl( a (a L_O + E_O (Q^2 - 2Mr)) - L_O \Delta \csc^2\theta \bigr) \\
&\qquad \times \bigl( a (a (L_A m_A + L_B m_B) + (E_A m_A + E_B m_B)(Q^2 - 2Mr)) \\
&\qquad - (L_A m_A + L_B m_B) \Delta \csc^2\theta \bigr) \sin^2\theta \; \bigl( (a^2+r^2)^2 - a^2 \Delta \sin^2\theta \bigr) \\
&\quad - a (Q^2 - 2Mr) \bigl( a (a L_O + E_O (Q^2 - 2Mr)) - L_O \Delta \csc^2\theta \bigr) \sin^2\theta \\
&\qquad \times \bigl( -a^4 (E_A m_A + E_B m_B) - 2a^2 (E_A m_A + E_B m_B) r^2 \\
&\qquad - (E_A m_A + E_B m_B) r^4 - a (L_A m_A + L_B m_B)(Q^2 - 2Mr) \\
&\qquad + a^2 (E_A m_A + E_B m_B) \Delta \sin^2\theta \bigr) \\
&\quad - a (Q^2 - 2Mr) \bigl( -a^4 E_O - 2a^2 E_O r^2 - E_O r^4 - a L_O (Q^2 - 2Mr) + a^2 E_O \Delta \sin^2\theta \bigr) \\
&\qquad \times \bigl( -(L_A m_A + L_B m_B) \Delta + a \bigl( a (L_A m_A + L_B m_B) \\
&\qquad + (E_A m_A + E_B m_B)(Q^2 - 2Mr) \bigr) \sin^2\theta \bigr) \\
&\quad + (\Delta - a^2 \sin^2\theta) \bigl( -a^4 (E_A m_A + E_B m_B) - 2a^2 (E_A m_A + E_B m_B) r^2 \\
&\qquad - (E_A m_A + E_B m_B) r^4 - a (L_A m_A + L_B m_B)(Q^2 - 2Mr) \\
&\qquad + a^2 (E_A m_A + E_B m_B) \Delta \sin^2\theta \bigr) \\
&\qquad \times \bigl( (a^2+r^2)(a L_O - E_O (a^2+r^2)) + a \Delta (-L_O + a E_O \sin^2\theta) \bigr) \Bigr].\label{16}
\end{aligned}
\end{equation}
From the conservation conditions \eqref{7} and the four-momentum expressions \eqref{15}, we can derive
\begin{equation}
\begin{aligned}
E_A m_A + E_B m_B &= E_O m_O,\quad L_A m_A + L_B m_B = L_O m_O,\\
m_A \sqrt{R_{A}} - m_B \sqrt{R_{B}} &= -m_O \sqrt{R_{O}},\quad m_A \sqrt{\Theta_{A}} - m_B \sqrt{\Theta_{B}} = -m_O \sqrt{\Theta_{O}}. \label{17}
\end{aligned}
\end{equation}
Substituting Eq.~\eqref{17} into Eq.~\eqref{16}, we obtain
\begin{equation}
\begin{aligned}
&-\,g_{\mu \nu}\left( m_A U_O^{\mu} U_A^{\nu} + m_B U_O^{\mu} U_B^{\nu} \right) = \frac{1}{\Delta^2 \rho^6} \Bigl[ \\
&\quad-\frac{1}{4} \Delta m_O R_O (a^2 + 2r^2 + a^2 \cos 2\theta)^2 
- \frac{1}{4} \Delta^2 m_O \Theta_O (a^2 + 2r^2 + a^2 \cos 2\theta)^2 \\
&\quad - \bigl( a (a L_O + E_O (Q^2 - 2Mr)) - L_O \Delta \csc^2\theta \bigr) \\
&\quad\times \bigl( a (a L_O m_O + E_O m_O (Q^2 - 2Mr)) - L_O m_O \Delta \csc^2\theta \bigr) \sin^2\theta \; \bigl( (a^2+r^2)^2 - a^2 \Delta \sin^2\theta \bigr) \\
&\quad - a (Q^2 - 2Mr) \bigl( a (a L_O + E_O (Q^2 - 2Mr)) - L_O \Delta \csc^2\theta \bigr) \sin^2\theta \\
&\qquad \times \bigl( -a^4 E_O m_O - 2a^2 E_O m_O r^2 - E_O m_O r^4 - a L_O m_O (Q^2 - 2Mr) + a^2 E_O m_O \Delta \sin^2\theta \bigr) \\
&\quad - a (Q^2 - 2Mr) \bigl( -a^4 E_O - 2a^2 E_O r^2 - E_O r^4 - a L_O (Q^2 - 2Mr) + a^2 E_O \Delta \sin^2\theta \bigr) \\
&\qquad \times \bigl( -L_O m_O \Delta + a \bigl( a L_O m_O + E_O m_O (Q^2 - 2Mr) \bigr) \sin^2\theta \bigr) \\
&\quad + (\Delta - a^2 \sin^2\theta) \bigl( -a^4 E_O m_O - 2a^2 E_O m_O r^2 - E_O m_O r^4 -\\
&\quad a L_O m_O (Q^2 - 2Mr) + a^2 E_O m_O \Delta \sin^2\theta \bigr) \\
&\qquad \times \bigl( (a^2+r^2)(a L_O - E_O (a^2+r^2)) + a \Delta (-L_O + a E_O \sin^2\theta) \bigr) \Bigr].\label{18}
\end{aligned}
\end{equation}
Consider the original expression
\begin{equation}
\mathcal{E} = \frac{1}{\Delta^2\rho^6}\Bigl[\,\cdots\,\Bigr],
\end{equation}
where the terms inside the square brackets are as given in Eq.~\eqref{18}.
Using the identity
\begin{equation}
a^2+2r^2+a^2\cos2\theta = 2(r^2+a^2\cos^2\theta)=2\rho^2,
\end{equation}
the first two terms become
\begin{equation}
-\frac14\Delta m_O R_O(2\rho^2)^2-\frac14\Delta^2 m_O\Theta_O(2\rho^2)^2
= -m_O\rho^4\bigl(\Delta R_O+\Delta^2\Theta_O\bigr).
\end{equation}
Define
\begin{equation}
\begin{aligned}
A &\equiv a\bigl(aL_O+E_O(Q^2-2Mr)\bigr)-L_O\Delta\csc^2\theta,\\
C &\equiv -E_O(a^2+r^2)^2-aL_O(Q^2-2Mr)+a^2E_O\Delta\sin^2\theta.
\end{aligned}  
\end{equation}
From the integral of motion $T_O=E_O(r^2+a^2)-L_O a$ and the four-velocity components
\begin{equation}
\rho^2U_O^t = -a(aE_O\sin^2\theta-L_O)+\frac{(r^2+a^2)T_O}{\Delta},\qquad
\rho^2U_O^\phi = -\Bigl(aE_O-\frac{L_O}{\sin^2\theta}\Bigr)+\frac{aT_O}{\Delta},
\end{equation}
we obtain
\begin{equation}
A = -\Delta\rho^2U_O^\phi,\qquad C = -\Delta\rho^2U_O^t.
\end{equation}
At the same time, it can be verified that the second bracket of the original sixth term also equals $C$.
After substituting $A$ and $C$, the last four terms become
\begin{equation}
\begin{aligned}
-\Delta^2\rho^4 m_O\Bigl[\sin^2\theta\bigl((a^2+r^2)^2-a^2\Delta\sin^2\theta\bigr)(U_O^\phi)^2 \\
+ 2a(Q^2-2Mr)\sin^2\theta\,U_O^\phi U_O^t - (\Delta-a^2\sin^2\theta)(U_O^t)^2\Bigr].   
\end{aligned}
\end{equation}
That is, the last four terms constitute a quadratic form in $U_O^t$ and $U_O^\phi$. From the geodesic equations
\begin{equation}
\rho^2U_O^r = \pm\sqrt{R_O},\qquad \rho^2U_O^\theta = \pm\sqrt{\Theta_O},
\end{equation}
we get
\begin{equation}
R_O = \rho^4(U_O^r)^2,\qquad \Theta_O = \rho^4(U_O^\theta)^2.
\end{equation}
Substituting all the terms into the original expression and factoring out the common factor, the original expression reduces to
\begin{equation}
m_O\Bigl[-\frac{\rho^2}{\Delta}(U_O^r)^2 - \rho^2(U_O^\theta)^2 - \frac{W}{\rho^2}\Bigr],
\end{equation}
where $W$ is precisely related to the $(t,\phi)$ part of the metric via
\begin{equation}
W = \rho^2\bigl(g_{tt}(U_O^t)^2 + 2g_{t\phi}U_O^tU_O^\phi + g_{\phi\phi}(U_O^\phi)^2\bigr) .
\end{equation}
Using $g_{rr}=\rho^2/\Delta,\; g_{\theta\theta}=\rho^2$ and the normalization of the four-velocity $g_{\mu\nu}U_O^\mu U_O^\nu = -1$, the entire square bracket finally equals $1$, so the original expression is identically equal to $m_O$. Thus, we have verified in Kerr--Newman spacetime the energy conservation relation given by Eq.~\eqref{8}, thereby obtaining
\begin{equation}
m_O = m_A\,\gamma_{O A} + m_B\,\gamma_{O B} = \sqrt{m_A^{2} + m_B^{2} + 2 m_A m_B \,\gamma_{A B}}. \label{30}
\end{equation}
This result reveals that the energy conservation in particle decay processes holds not only in Kerr black holes \cite{5} and Kerr--de Sitter black holes \cite{6}, but also in the Kerr--Newman background.

\section{Numerical results}
\begin{figure}[!h]
  \centering
    \includegraphics[width=0.5\linewidth]{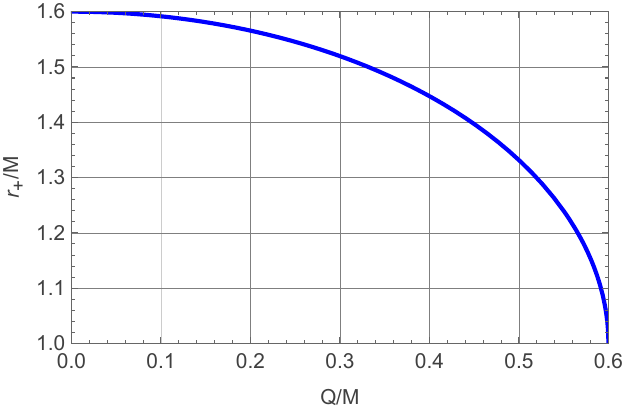} 
  \caption{Dimensionless event horizon $r_+/M$ versus dimensionless charge $Q/M$ for $a/M=0.8$.}
\label{fig:1}
\end{figure}

In this section, we verify the energy conservation property in the particle decay process in Kerr--Newman spacetime through concrete numerical calculations, and compare it with the case of a Kerr black hole. For the treatment of the particle decay process, we follow the particle splitting procedure of Refs.\ \cite{7,8,9,10,11,19,20}. In those references, repeated Penrose processes are studied, while for the particle splitting process it suffices to consider the first Penrose process.
We consider particle motion in the equatorial plane. Particle $O$ splits into particles $A$ and $B$, and the splitting formulas satisfy Eqs.\ (18) and (19) of Ref.\ \cite{9}. Since the presence of charge reduces the spin value at which the black hole becomes extremal, for the black hole spin we set $a=0.8M$, and the charge $Q$ then takes values in the range $0$--$0.6M$.
The black hole horizons satisfy
\begin{equation}
r_{\pm} = M \pm \sqrt{M^2 - a^2 - Q^2}.
\end{equation}
The decay radius must be larger than the outer horizon to be physically meaningful. In Fig.\ \ref{fig:1}, we plot the dimensionless event horizon $r_+/M$ as a function of the dimensionless charge $Q/M$ for $a=0.8M$, which provides a reference for choosing the decay radius at different charge values.
For the particles involved in the splitting, we adopt the parameters from \cite{5,7,9}:
\begin{equation}
\hat{E}_O = 1,\quad \hat{\phi}_A = -19.434,\quad \frac{\mu_B}{\mu_A} = 0.78345.   
\end{equation}
The meaning of the above symbols can be found in \cite{5,7,9}, with the only difference being that our $O, A, B$ correspond to their $0, 1, 2$, respectively.

We first verify energy conservation for the case $Q/M=0.4$. Of course, as demonstrated in Refs.\ \cite{5,6}, the numerical precision must be sufficiently high. We take the decay radii as $r/M=1.7, 1.8, 1.9$.
\begin{table}[htbp]
\centering
\caption{Recovered mass and conserved quantity evaluated at three different decay radii, for $Q/M=0.4$.}
\label{tab:1}
\begin{tabular}{c c c c c}
\toprule
$r/M$ & Particle & $m_i/m_O$ & $\hat{E_i}$ & $\hat{\phi_i}$ \\
\midrule
\multirow{3}{*}{1.70}
& O & 1 & 1 & 2.697541625762404 \\
& A & 0.01788385302375085 & -1.625862462085866  & -19.434 \\
& B & 0.0140111046514576 & 73.4472199664657  & 217.3345001108953 \\
\cmidrule{2-5}
\multirow{3}{*}{1.80}
& O & 1 & 1 & 2.687439320554617 \\
& A & 0.01910638676144957 & -0.7760197731433452  & -19.434 \\
& B & 0.01496889870825766 & 67.79569784652061  & 204.3405397077911 \\
\cmidrule{2-5}
\multirow{3}{*}{1.90}
& O & 1   & 1 & 2.693624376449537 \\
& A & 0.02031088856703504  & -0.0816924618964823   & -19.434 \\
& B & 0.0159125656478436 & 62.94768981054206  & 194.0822274175387 \\
\bottomrule
\end{tabular}
\end{table}
\begin{table}[htbp]
\centering
\caption{$Q/M=0.4$, recovered four-velocity $U^\mu$ and normalization check $U^\mu U_\mu$.}
\label{tab:2}
\begin{tabular}{c c c c}
\toprule
$r/M$ & Particle & $U^\mu = (U^t, U^r, U^\theta, U^\phi)$ & $U^\mu U_\mu$ \\
\midrule
\multirow{3}{*}{1.70}
& O & (6.303868399980724, 0, 0, 1.966185933639372) & -1.000000000000004 \\
& A & (36.29039317403336, 0, 0, 3.087536688072346) & -0.999999999999936\\
& B & (403.5981803685903, 0, 0, 136.3895944568258) & -0.999999995197868\\
\cmidrule{2-4}
\multirow{3}{*}{1.80}
& O & (5.174640144383902, 0, 0, 1.553389545376737) & -1.000000000000002\\
& A & (29.47417444957069, 0, 0, 1.228390561384338) & -0.999999999999994\\
& B & (308.0717731981189, 0, 0, 102.2065463889355) & -0.999999997024133\\
\cmidrule{2-4}
\multirow{3}{*}{1.90}
& O & (4.46313328903272, 0, 0, 1.285677884158991) & -1\\
& A & (25.04342702153362, 0, 0, 0.1567283733515471) & -0.999999999999996\\
& B & (248.5129752786012, 0, 0, 80.5963425392908) & -0.999999999814463\\
\bottomrule
\end{tabular}
\end{table}
The last column of Table~\ref{tab:2} reflects that, to high precision, the normalization condition holds.
\begin{table}[htbp]
\centering
\caption{Relative Lorentz factors $\gamma_{OA}$, $\gamma_{OB}$ and $\gamma_{AB}$ ($Q/M=0.4$).}
\label{tab:3}
\begin{tabular}{c c c c}
\toprule
$r/M$ & $\gamma_{OA}$ & $\gamma_{OB}$ & $\gamma_{AB}$ \\
\midrule
1.70 & 27.96163443688963 & 35.68157200044971 & 1994.4002474465 \\
1.80 & 26.17294935390826 & 33.39788161440413 & 1747.2122349735 \\
1.90 & 24.6212596545926 & 31.41670236209086 & 1546.00768414485 \\
\bottomrule
\end{tabular}
\end{table}
\begin{table}[htbp]
\centering
\caption{Verification of energy conservation ($Q/M=0.4$).}
\label{tab:4}
\begin{tabular}{c c c}
\toprule
$r/M$ & $m_A\gamma_{OA}+m_B\gamma_{OB}$ & $\sqrt{m_A^2+m_B^2+2m_Am_B\gamma_{AB}}$ \\
\midrule
1.70 & 1.000000000000005 $m_O$ & 1.000000000000474 $m_O$\\
1.80 & 0.999999999999999 $m_O$& 1.000000000000333 $m_O$\\
1.90 &  $m_O$& 1.000000000000023 $m_O$\\
\bottomrule
\end{tabular}
\end{table}
The results in Table~\ref{tab:4} convincingly illustrate that energy is conserved in Kerr–Newman spacetime. Furthermore, from Tables~\ref{tab:1} to \ref{tab:4}, it can be seen that the larger the decay radius, the larger the masses of particles $A$ and $B$, the smaller the absolute values of their specific energies; the smaller the three relative Lorentz factors; and the smaller the four-velocities of the three particles. These phenomena are completely consistent with the patterns in Ref.~\cite{6}.

Next, we will analyze the effect of charge on the particle splitting process. For the decay radius, we take $r/M=1.8$.
\begin{figure}[!h]
  \centering
  \begin{subfigure}{0.45\textwidth}
    \centering
    \includegraphics[width=\linewidth]{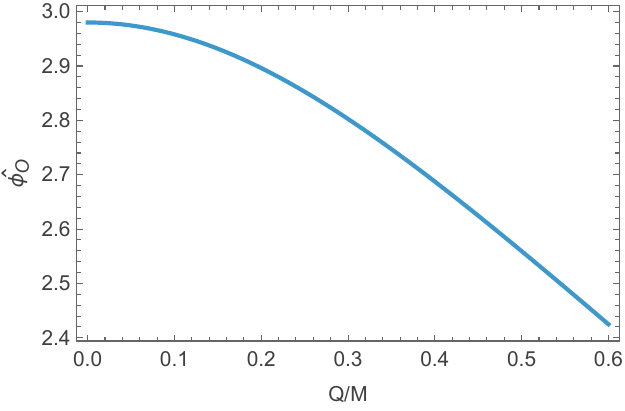}
  \end{subfigure}
  \begin{subfigure}{0.45\textwidth}
    \centering
    \includegraphics[width=\linewidth]{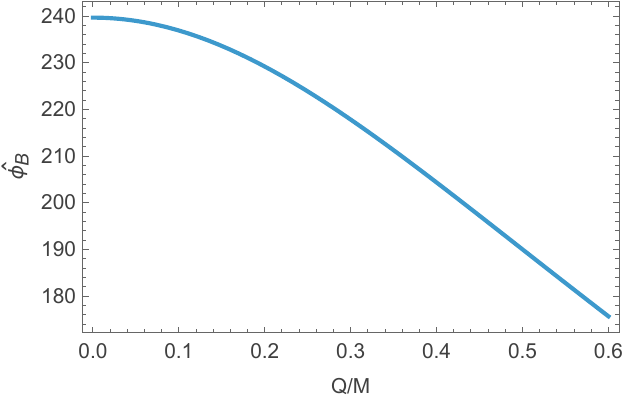}  
  \end{subfigure}
\caption{Variation of the specific angular momenta of the particles with charge.}
  \label{fig:2}
\end{figure}
\begin{figure}[!h]
  \centering
  \begin{subfigure}{0.45\textwidth}
    \centering
    \includegraphics[width=\linewidth]{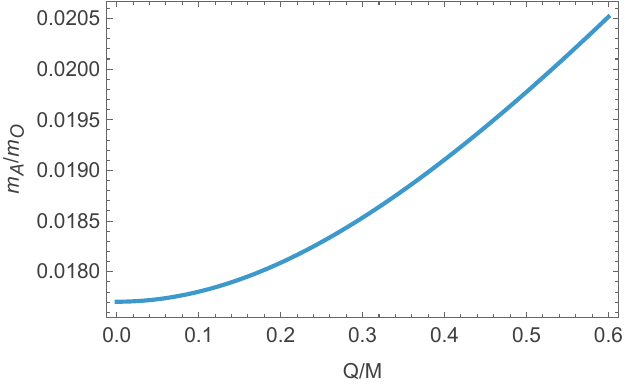}
  \end{subfigure}
  \begin{subfigure}{0.45\textwidth}
    \centering
    \includegraphics[width=\linewidth]{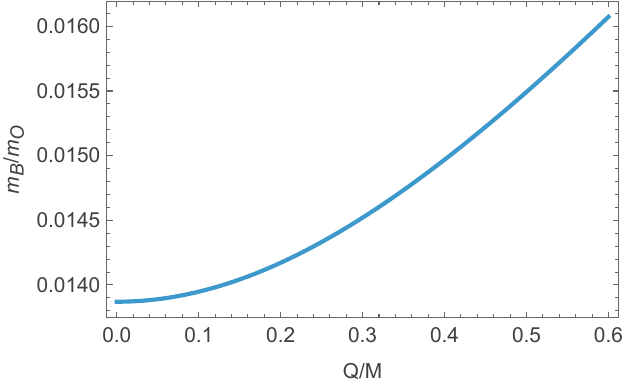}  
  \end{subfigure}
\caption{Variation of the masses of the particles with charge.}
  \label{fig:3}
\end{figure}
\begin{figure}[!h]
  \centering
  \begin{subfigure}{0.45\textwidth}
    \centering
    \includegraphics[width=\linewidth]{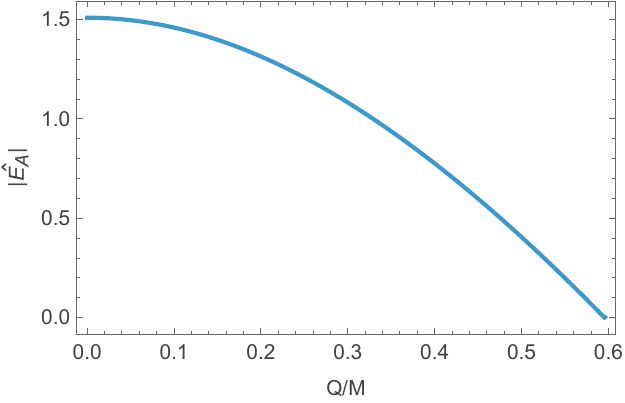}
  \end{subfigure}
  \begin{subfigure}{0.45\textwidth}
    \centering
    \includegraphics[width=\linewidth]{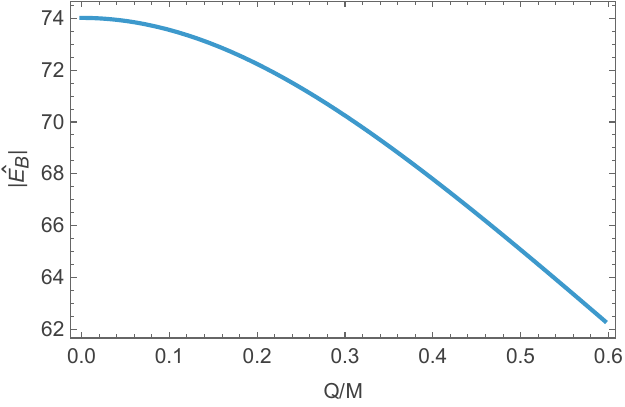}  
  \end{subfigure}
\caption{Variation of the absolute values of the specific energies of the particles with charge.}
  \label{fig:4}
\end{figure}
\begin{figure}[!h]
  \centering
    \includegraphics[width=\linewidth]{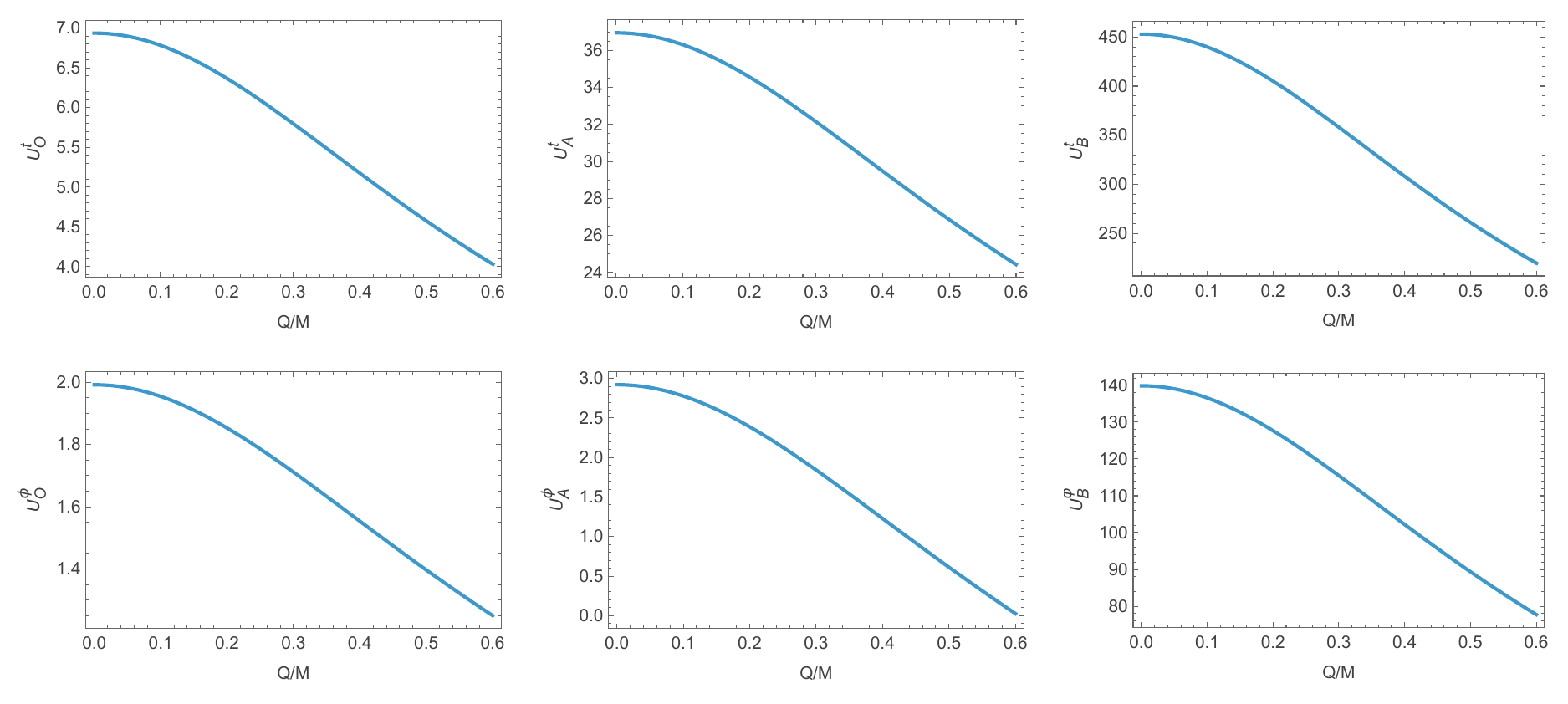} 
  \caption{Variation of the four-velocities of the particles with charge.}
\label{fig:5}
\end{figure}
\begin{figure}[!h]
  \centering
  \begin{subfigure}{0.32\textwidth}
    \centering
    \includegraphics[width=\linewidth]{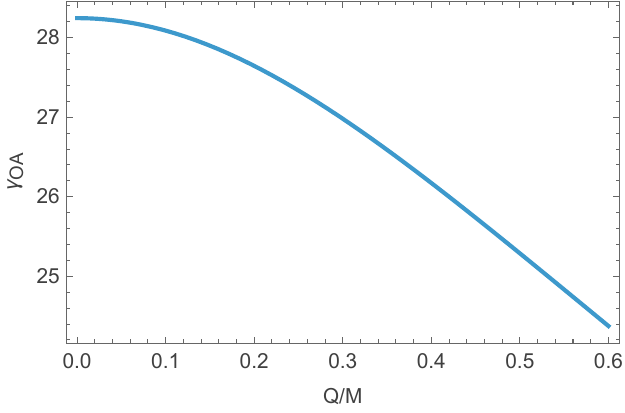}
  \end{subfigure}
  \begin{subfigure}{0.32\textwidth}
    \centering
    \includegraphics[width=\linewidth]{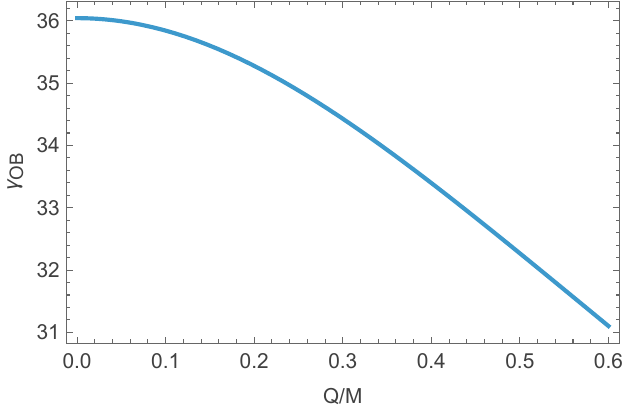}  
  \end{subfigure}
  \begin{subfigure}{0.32\textwidth}
    \centering
    \includegraphics[width=\linewidth]{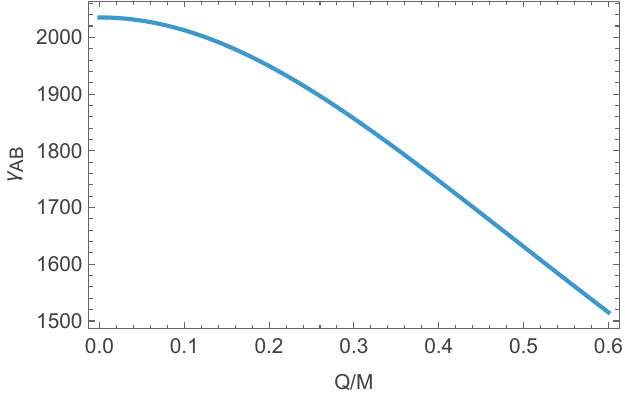}  
  \end{subfigure}
\caption{Variation of the relative Lorentz factors of the particles with charge.}
  \label{fig:6}
\end{figure}
From Figs.\ \ref{fig:2} to \ref{fig:6}, it is clearly seen that as the charge increases, the specific angular momenta of the parent particle $O$ and the daughter particle $B$ become smaller; the masses of the daughter particles $A$ and $B$ become larger, and the absolute values of their specific energies become smaller; the relative Lorentz factors among all three particles become smaller; and the four-velocities of all three particles become smaller. These phenomena indicate that the charge has an opposite trend compared with the cosmological constant in Kerr--de Sitter spacetime \cite{6}. Moreover, except for the special case of the specific angular momenta, the pattern of variation with a larger decay radius is similar to that with a larger charge.

\section{Conclusion}
In this paper, we have proved that a mass deficit exists in the particle decay process within the Kerr--Newman black hole, and this deficit is transformed into kinetic energy in the center-of-mass frame. This is exactly what Eq.~\eqref{30} demonstrates. It implies that the sum of the rest masses of the decay products must be less than the mass of the parent particle, unless there is never any relative velocity between the two daughter particles or they never separate from each other---a situation that cannot actually occur. This is consistent with the cases of the Kerr black hole and the Kerr--de Sitter black hole. Denoting the parent particle as $O$ and the daughter particles as $A$ and $B$, the numerical results not only further confirm the correctness of the energy conservation, but also reveal the characteristics of particle splitting in Kerr--Newman spacetime: namely, the larger the charge parameter $Q$, the smaller the specific angular momenta of the parent particle $O$ and the daughter particle $B$; the larger the masses of the daughter particles $A$ and $B$ but the smaller the absolute values of their specific energies; the smaller the relative Lorentz factors among the three particles; and the smaller their four-velocities. These trends are opposite to those for increasing the cosmological constant in Kerr--de Sitter spacetime, and are the same as those for increasing the decay radius.
On the other hand, the energy conservation relation in Kerr--Newman spacetime provides a concise local criterion for energy extraction via the Penrose process in this background.

\noindent {\bf Acknowledgments}

\noindent
This work is supported by the National Natural Science Foundation of China (Grants Nos.
12375043, 12575069 ), and Chongqing Normal University Fund Project (Grants No. 26XLB001).

\end{document}